# Georges Lemaître and Stigler's Law of Eponymy

**David L. Block**

**Abstract** One of the greatest discoveries of modern times is that of the expanding Universe, almost invariably attributed to Hubble (Proceedings of the National Academy of Sciences of the United States of America 15:168, 1929). What is not widely known is that the original treatise by Lemaître (Annales de la Société Scientifique de Bruxelles, Sèrie A 47:49, 1927) contained a rich fusion of both theory and of observation. The French paper was meticulously censored when published in English: all discussions of radial velocities and distances, and the very first empirical determination of $H_0$, were suppressed. Stigler's law of eponymy is yet again affirmed: no scientific discovery is named after its original discoverer (Merton, American Sociological Review 22(6):635, 1957). An appeal is made for a Lemaître Telescope naming opportunity, to honour the discoverer of the expanding universe.

## Lemaître (1927): A Theoretical Paper?

The title of the original 1927 paper indicates to the reader that the content will be a fusion of both theory *and of observation*: "Un univers homogène de masse constante et de rayon croissant, rendant compte de la vitesse radiale des nébuleuses extra-galactiques." Which translates into English thus: "A homogeneous universe of constant mass and increasing radius accounting for the radial velocity of extra-galactic nebulae".

Lemaître spent the years 1924–1925 at the Harvard College Observatory. He had an excellent foundation in observational astronomy, writing about terms such as the effective temperatures of stars, trigonometric parallaxes, moving-cluster parallaxes, absolute bolometric magnitudes, dwarf branch stars, giant branch stars, and the like.

D.L. Block (✉)
School of Computational and Applied Mathematics, University of the Witwatersrand, Johannesburg, South Africa
e-mail: david.block@wits.ac.za

To speak of Lemaître (1927) as a most remarkable and absolutely brilliant theoretical paper only, is a grave injustice to the very title. Not only does Lemaître derive a linear relationship between the radial velocities of galaxies and their distances in the above paper, but he is eager to determine the rate at which the universe expands. Lemaître (1927) carefully uses the radial velocities of 42 extra-galactic nebulae tabulated by Strömberg (1925), and he converts apparent magnitudes $m$ into distance [$\log r = 0.2\ m + 4.04$] following Hubble (1926). The actual value which Lemaître obtains in 1927 for the rate of expansion of the Universe is 625 km s$^{-1}$ Mpc$^{-1}$; 575 km s$^{-1}$ Mpc$^{-1}$ with different weighting factors (Fig. 1).

Jaki (1974) elaborates: "Lemaître's treatment of the problem could hardly be more impressive with respect to specific results . . . a formula and a table of values for the redshift of receding galaxies in fine agreement with the actually observed data . . ."

When the Royal Astronomical Society decided to publish an English translation in 1931 from the journal Annales de la Société Scientifique de Bruxelles, a most dramatic censorship of the first empirical determination of $H_0$ occurred (Fig. 2). A meticulously researched book (with a foreword by the late Allan Sandage) has been published on this precise theme. It is entitled *Discovering the Expanding Universe* (Nussbaumer and Bieri 2009). Professor Nussbaumer graciously sent me a copy of the original French paper in 2009, and the sectors censored out in the English translation appear in Fig. 2. Equation (24) holds the key. In an independent study, Sidney van den Bergh (2011) affirms that the suppressions in Eq. (24) were intentional.

It would be historically accurate to say that the *testing* of a linear velocity-distance relation is due to the meticulous observations by Hubble and Humason in subsequent years, but not the *formulation* of this relation, as seen in the complete original equation (24).

## Priorities in Scientific Discovery

And now, I give some insight into the mindset of Edwin Hubble. He was fiercely territorial, as we see in a letter from Hubble to de Sitter, dated 21 August 1930, wherein Hubble writes: "I consider the velocity-distance relation, its *formulation*, testing and confirmation, as a Mount Wilson contribution and I am deeply concerned in its recognition as such" (emphasis added).

Nussbaumer and Bieri (2009) respond as follows:

> . . . the *formulation* and its central place in cosmology was first given by Lemaître . . . there is no justification to glorify Hubble's publication of 1929 [as the] original discovery of the linear velocity- distance relationship . . . (emphasis, mine).

Lemaître was eclipsed. Multitudes of textbooks proclaim Hubble as the discoverer of the expanding universe. But herein lies a repeated pattern. In 1927,

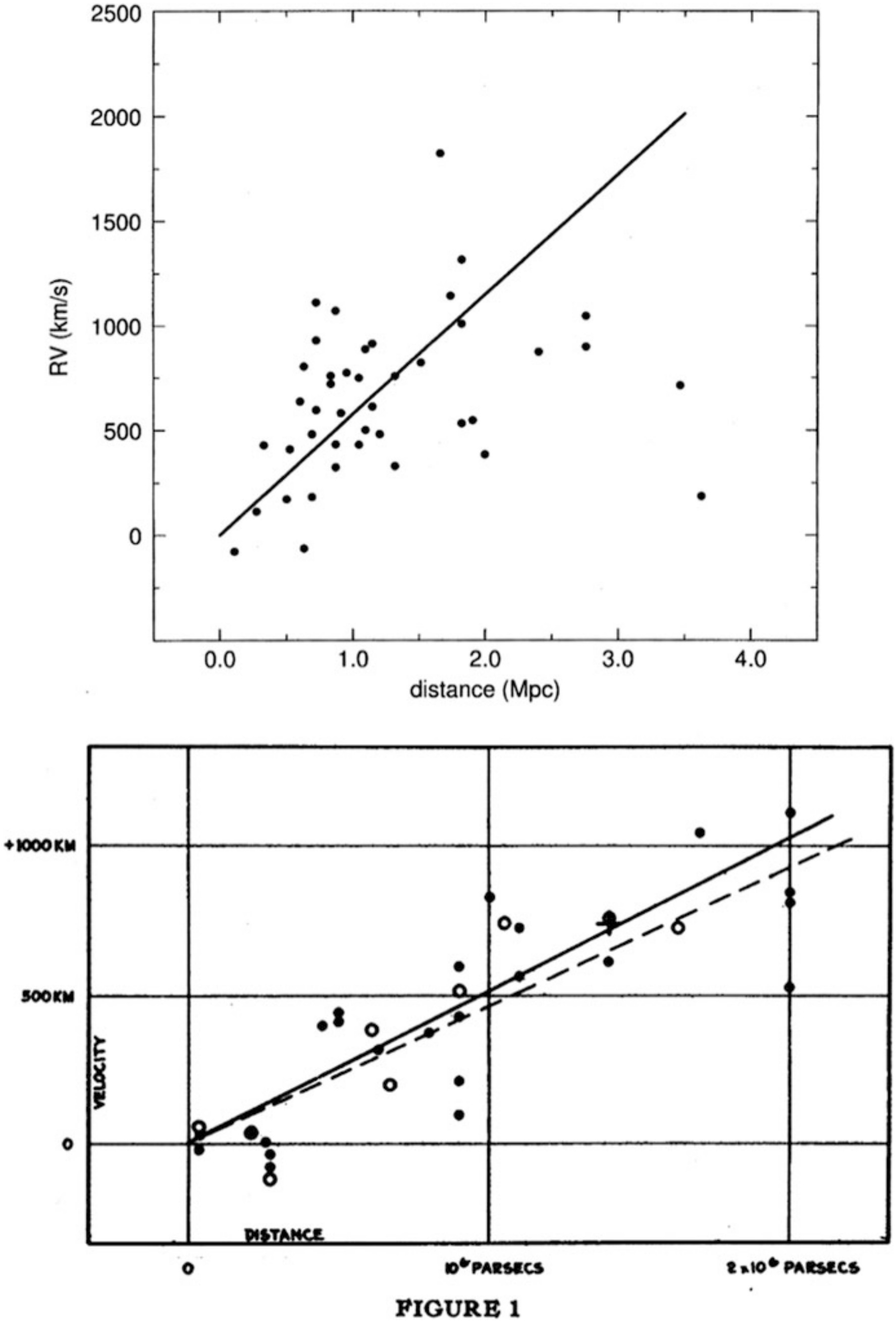

**Fig. 1** *Upper panel*: The data used by Lemaître (1927) to yield the first empirical value of the rate of expansion of the Universe in which *v*/*r* is predicted to be constant (see Eq. 24 in Fig. 2). Lemaître derived values of 625 km $s^{-1}$ $Mpc^{-1}$ and 575 km $s^{-1}$ $Mpc^{-1}$. The *solid line in the top panel* has a slope of 575 km $s^{-1}$ $Mpc^{-1}$ and is reconstructed by H. Duerbeck. *Lower panel*: the radial velocity–distance diagram published by Hubble, 2 years later, in 1929, with best slope of 530 km $s^{-1}$ $Mpc^{-1}$ (*Top panel*: Courtesy H. Duerbeck)

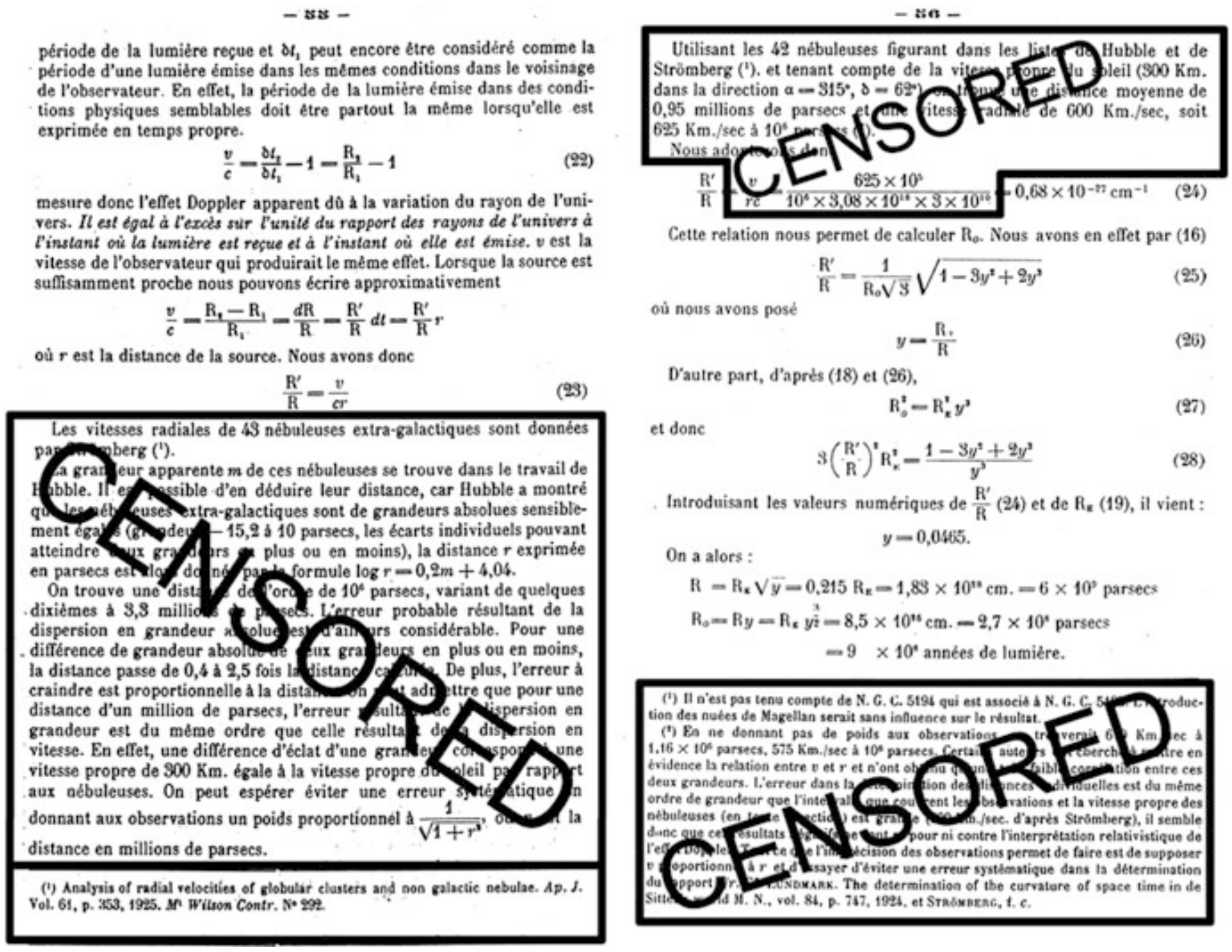
— 55 —

période de la lumière reçue et $\delta t_1$ peut encore être considéré comme la période d'une lumière émise dans les mêmes conditions dans le voisinage de l'observateur. En effet, la période de la lumière émise dans des conditions physiques semblables doit être partout la même lorsqu'elle est exprimée en temps propre.

$$\frac{v}{c} = \frac{\delta t_2}{\delta t_1} - 1 = \frac{R_2}{R_1} - 1 \qquad (22)$$

mesure donc l'effet Doppler apparent dû à la variation du rayon de l'univers. *Il est égal à l'excès sur l'unité du rapport des rayons de l'univers à l'instant où la lumière est reçue et à l'instant où elle est émise.* $v$ est la vitesse de l'observateur qui produirait le même effet. Lorsque la source est suffisamment proche nous pouvons écrire approximativement

$$\frac{v}{c} = \frac{R_2 - R_1}{R_1} = \frac{dR}{R} = \frac{R'}{R}\,dt = \frac{R'}{R}\,r$$

où $r$ est la distance de la source. Nous avons donc

$$\frac{R'}{R} = \frac{v}{cr} \qquad (23)$$

Les vitesses radiales de 43 nébuleuses extra-galactiques sont données pa[illegible]berg (¹).

[illegible] grandeur apparente $m$ de ces nébuleuses se trouve dans le travail de [illegible]bble. Il [illegible]ssible d'en déduire leur distance, car Hubble a montré qu[illegible] extra-galactiques sont de grandeurs absolues sensiblement éga[illegible] — 15,2 à 10 parsecs, les écarts individuels pouvant atteindre [illegible] plus ou en moins), la distance $r$ exprimée en parsecs est [illegible] formule $\log r = 0{,}2m + 4{,}04$.

On trouve une dista[illegible] de $10^6$ parsecs, variant de quelques dixièmes à 3,3 millio[illegible]secs. L'erreur probable résultant de la dispersion en grandeur [illegible] considérable. Pour une différence de grandeur absol[illegible] grandeurs en plus ou en moins, la distance passe de 0,4 à 2,5 fois la distance [illegible]. De plus, l'erreur à craindre est proportionnelle à la dista[illegible] admettre que pour une distance d'un million de parsecs, l'erreur [illegible] dispersion en grandeur est du même ordre que celle résulta[illegible] dispersion en vitesse. En effet, une différence d'éclat d'une gran[illegible] une vitesse propre de 300 Km. égale à la vitesse propre d[illegible]leil pa[illegible] rapp[illegible]t aux nébuleuses. On peut espérer éviter une erreur [illegible]atique [illegible] donnant aux observations un poids proportionnel à $\frac{1}{\sqrt{1+r^2}}$, [illegible] la distance en millions de parsecs.

(¹) Analysis of radial velocities of globular clusters and non galactic nebulae. *Ap. J.* Vol. 61, p. 353, 1925. *Mt Wilson Contr.* N° 292.

— 56 —

Utilisant les 42 nébuleuses figurant dans les liste[illegible] Hubble et de Strömberg (¹). et tenant compte de la vi[illegible]leil (300 Km. dans la direction $\alpha = 315°$, $\delta = 62°$) [illegible] distance moyenne de 0,95 millions de parsecs [illegible] de 600 Km./sec, soit 625 Km./sec à $10^6$ [illegible].

Nous ado[illegible]

$$\frac{R'}{R} = \frac{v}{rc} = \frac{625 \times 10^5}{10^6 \times 3{,}08 \times 10^{18} \times 3 \times 10^{10}} = 0{,}68 \times 10^{-27}\ \mathrm{cm}^{-1} \qquad (24)$$

Cette relation nous permet de calculer $R_0$. Nous avons en effet par (16)

$$\frac{R'}{R} = \frac{1}{R_0\sqrt{3}}\sqrt{1 - 3y^2 + 2y^3} \qquad (25)$$

où nous avons posé

$$y = \frac{R_0}{R} \qquad (26)$$

D'autre part, d'après (18) et (26),

$$R_0^2 = R_E^2\, y^3 \qquad (27)$$

et donc

$$3\left(\frac{R'}{R}\right)^2 R_E^2 = \frac{1 - 3y^2 + 2y^3}{y^3} \qquad (28)$$

Introduisant les valeurs numériques de $\frac{R'}{R}$ (24) et de $R_E$ (19), il vient :

$$y = 0{,}0465.$$

On a alors :

$$R = R_E\sqrt{y} = 0{,}215\ R_E = 1{,}83 \times 10^{26}\ \mathrm{cm.} = 6 \times 10^7\ \mathrm{parsecs}$$

$$R_0 = Ry = R_E\, y^{\frac{3}{2}} = 8{,}5 \times 10^{24}\ \mathrm{cm.} = 2{,}7 \times 10^6\ \mathrm{parsecs}$$

$$= 9 \times 10^6\ \text{années de lumière.}$$

(¹) Il n'est pas tenu compte de N. G. C. 5194 qui est associé à N. G. C. 51[illegible]roduction des nuées de Magellan serait sans influence sur le résultat.

(²) En ne donnant pas de poids aux observations [illegible] trouverait 6[illegible] Km[illegible]ec à 1,16 × 10⁶ parsecs, 575 Km./sec à 10⁶ parsecs. Certai[illegible] aute[illegible] cherch[illegible] en évidence la relation entre $v$ et $r$ et n'ont ob[illegible] faibl[illegible] corr[illegible]tion entre ces deux grandeurs. L'erreur dans la [illegible] des [illegible] individuelles est du même ordre de grandeur que l'inte[illegible] les observations et la vitesse propre des nébuleuses (en [illegible] est gra[illegible]m./sec. d'après Strömberg), il semble donc que ce[illegible] pour ni contre l'interprétation relativistique de l'ef[illegible] l'im[illegible]écision des observations permet de faire est de supposer $v$ [illegible]oportionn[illegible] à $r$ et d'[illegible]ssayer d'éviter une erreur systématique dans la détermination du [illegible]pport [illegible]UNDMARK. The determination of the curvature of space time in de Sitte[illegible]d M. N., vol. 84, p. 747, 1924, et STRÖMBERG, *l. c.*

**Fig. 2** Sections in *black boxes*, pertaining to the discussion and use of radial velocities of galaxies and their distances by Lemaître (1927) to provide the first empirical determination of $H_0$ were meticulously and ingeniously suppressed or censored in the English translation. Equation (24) is absolutely crucial

Knut Lundmark penned these words, cited by Sandage (2004): "As to Hubble's way of acknowledging his predecessors I have no reason to enter upon this question here."

Is it not strange that Vesto Slipher is not referenced *at all* in Hubble's landmark paper of 1929? The vast majority of radial velocities in that paper are from Slipher. Perhaps an even more glaring example is Fig. 3, written to J. H. Reynolds on a visit to England.

As elucidated by Block and Freeman (2008), Reynolds rises to the Hubble request. He publishes his results in Reynolds (1920). Hubble very carefully studied this paper and actually pencilled in some handwritten comments, shown to me by the late Allan Sandage. (For example, next to each of the Reynolds class II, III and IV are the Sa, Sb and Sc notations penciled in by Hubble. Dr. Sandage furthermore affirmed to me that the correspondence between Reynolds types and Hubble types is "one-to-one"). Hubble (1926) appeared in print 6 years after Reynolds – with no reference to Reynolds (1920). Was Lundmark correct?

In the English speaking world, a total eclipse fell on the remarkable astronomical insight of Lemaître (Kragh and Smith 2003). The translator has been demonstrated to be Lemaître himself (see below my Note Added in Proof). What an intriguing

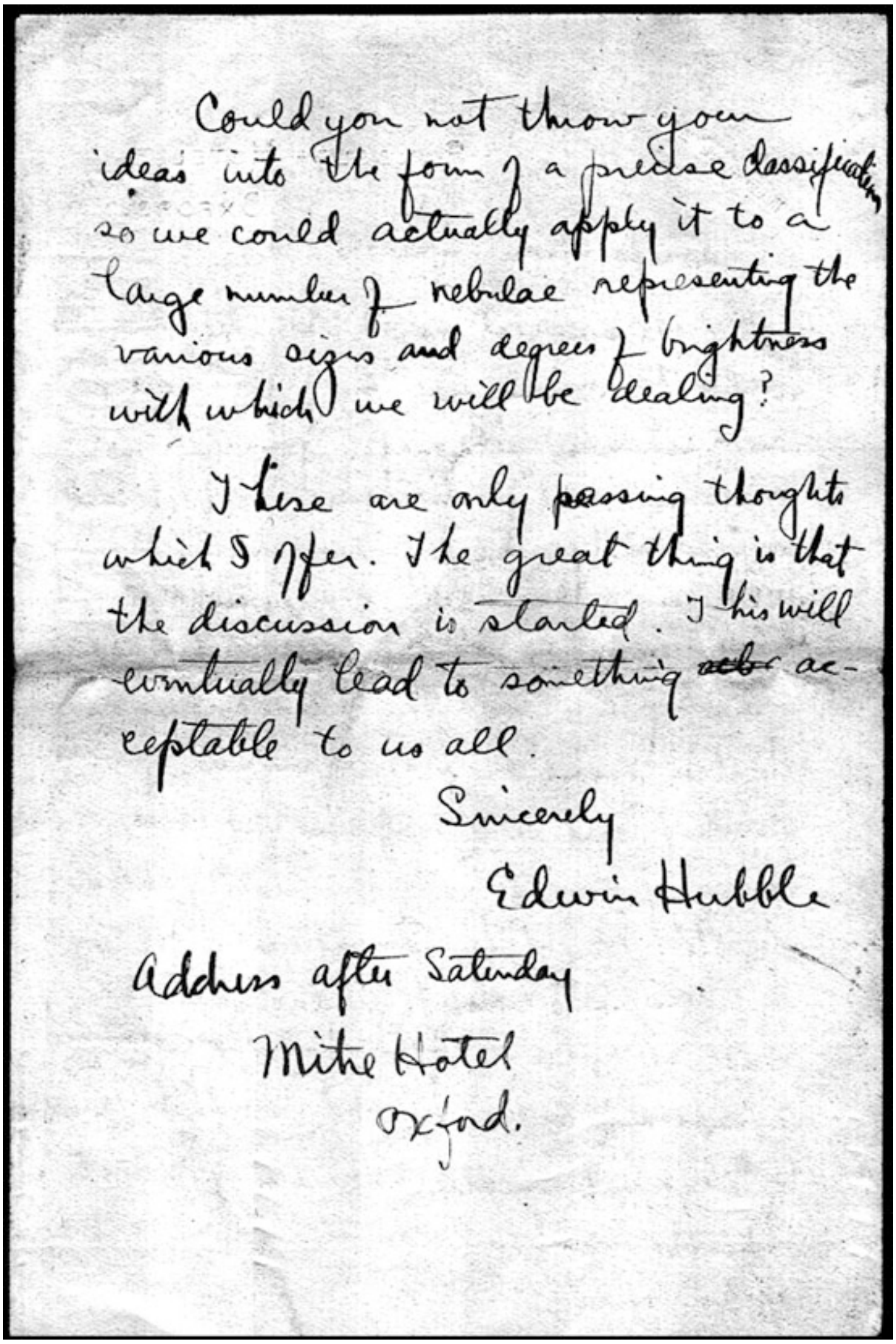

Could you not throw your ideas into the form of a precise classification so we could actually apply it to a large number of nebulae representing the various sizes and degrees of brightness with which we will be dealing?

These are only passing thoughts which I offer. The great thing is that the discussion is started. This will eventually lead to something ~~acc~~ acceptable to us all.

Sincerely
Edwin Hubble

Address after Saturday
Mitre Hotel
Oxford.

**Fig. 3** Hubble requests the following from J. H. Reynolds: "Could you not throw your ideas into the form of a precise classification so we could actually apply it to a large number of nebulae representing the various sizes and degrees of brightness with which we will be dealing?" The letter is believed to have been written in 1919, a year in which Hubble is recorded to have dined in England. This letter was first reproduced in Block and Freeman (2008). The original is in the archives of the Royal Astronomical Society of London

proof of Stigler's Law of Eponymy; Lemaître was, through his own actions, robbed of being attributed with one of the greatest discoveries in astronomy of all time. There are myriads of speculations as to why Lemaître decided to omit his empirical computation of the rate of expansion of the universe from the English

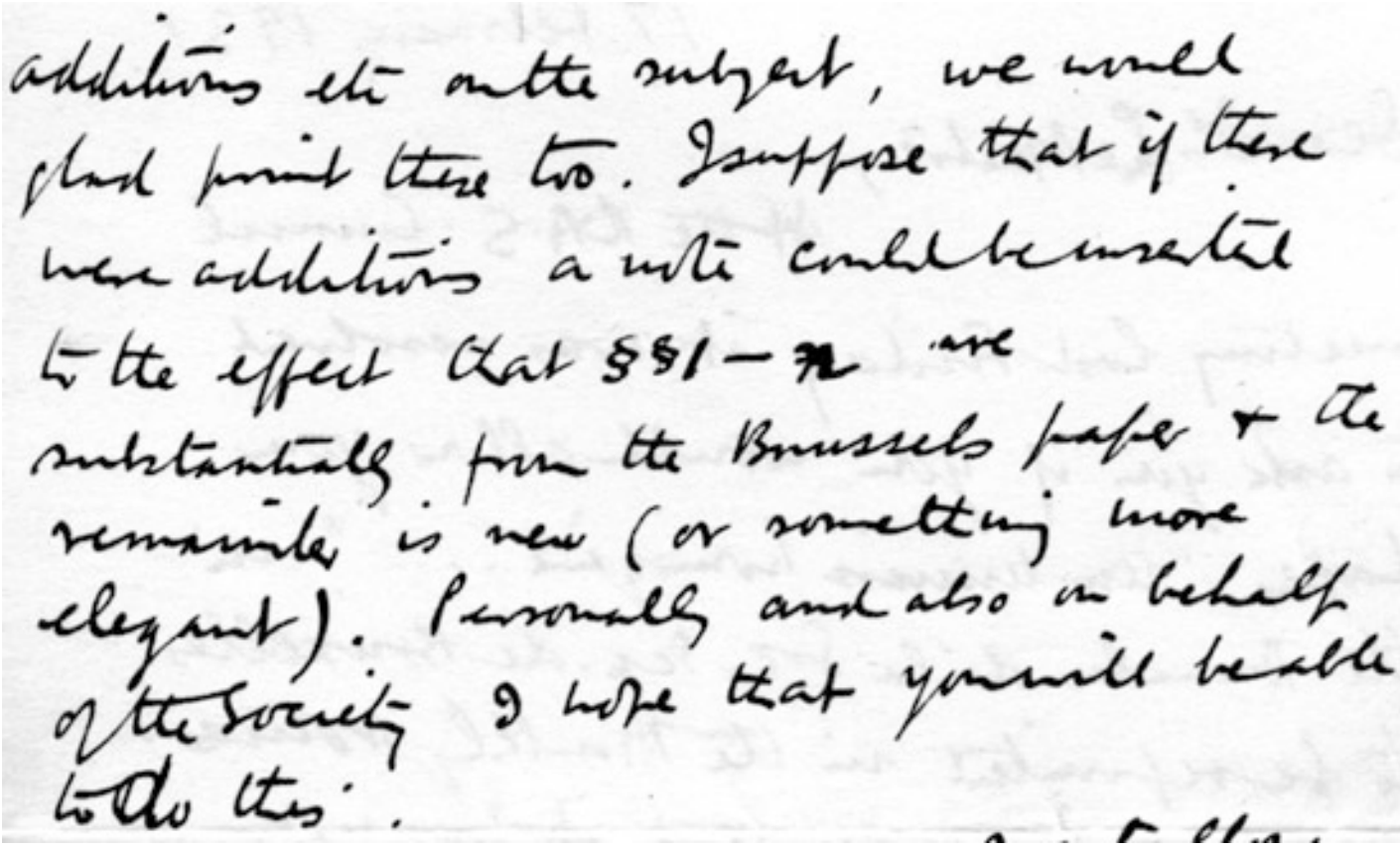
additions etc on the subject, we would
gladly print these too. I suppose that if there
were additions a note could be inserted
to the effect that §§1–72 are
substantially from the Brussels paper + the
remainder is new (or something more
elegant). Personally and also on behalf
of the Society I hope that you will be able
to do this.

**Fig. 4** The alarming "presence of a censor" is seen in this February 1931 letter from WM Smart to G. Lemaître. In extremely polite terms, Lemaître is told by Smart that Hubble's observational result of 1929 is "something more elegant". The reason we know that Smart is specifically alluding to Hubble (1929) is as follows: Lemaître is given full freedom to translate his 1927 French paper, from paragraph 1 to paragraph 72 (which at first glance, appears as a symbol "n", but which is actually the number "72" as affirmed by D. Lambert – private communication). Here follows the punch-line: paragraph 73 is Lemaître's equation 24. Paragraph 73 would have been the empirical determination by Lemaître of his expansion coefficient, published in 1927 (Courtesy: Lemaître Archives, Louvain-la-Neuve)

translation of his monumental 1927 paper (Fig. 4), although historians of astronomy must never forget his *original* intentions, as recalled by Lemaître himself several years later (in 1950, see below). The history is not irrelevant.

## CODA: A Lemaître ELT?

One of Galileo's masterful works was entitled *Sidereus Nuncius* – the starry messenger. The moral of the censorship (Fig. 2) is – as Martin Gaskell (private communication) poignantly reminded me – Mark chapter 4, verse 22. I allow Nussbaumer and Bieri (2009) to have the final word here regarding the legendary Georges Lemaître: "Even in his influential *The Realm of the Nebulae* published in 1936, he [Hubble] avoided any reference to Lemaître. Was he afraid that a gem might fall from his crown if people became aware of Lemaître's pioneering *fusion of observation and theory* 2 years before Hubble delivered the *confirmation*?" (italics, mine).

**Acknowledgements** First and foremost, I thank my co-author of *Shrouds of the Night*, K. C. Freeman, for his invaluable insight, encouragement and support. I am indebted to Harry Nussbaumer, Robert Smith and Sidney van den Bergh for their detailed comments on the manuscript. I warmly thank Dominique Lambert, Piet vd Kruit, Maarten Schmidt and the Director General of ESO, Tim de Zeeuw for their insight and interest. Profound appreciation goes to my sponsors AVENG and AECI for financial support and to archivist Mrs. Liliane Moens at the Lemaître Archives.

## Note Added in Proof – "The History of This Science Competition Is Not Irrelevant" – Reflections by Lemaître Himself, in 1950

The world has before its eyes one of the most brilliant examples of *Stigler's law of eponymy* –which in its simplest form, asserts that: no scientific discovery is named after its original discoverer. "Priorities in Scientific Discovery: A Chapter in the Sociology of Science" (Merton 1957) is of crucial importance in this context.

In a Comment published in *Nature* Mario Livio (*Nature*, 479, 171, 2011) has unearthed a letter from Lemaître to W. M. Smart (dated 9 March 1931). From that document, it is clear that Lemaître himself translated his 1927 paper into English and who also omitted his determination of the coefficient of expansion of the Universe ($H_0$) from values of radial velocities available as of 1927. However, in his Comment Livio omits a vital reference, namely thoughts penned by Lemaître himself in 1950 (*L'expansion de l'Univers, Bibliographie: Annales d'Astrophysique*, 13, 344):

> About my contribution of 1927, I do not want to discuss if I was a professional astronomer. I was, in any event, an IAU member (Cambridge, 1925), and I had studied astronomy for two years, a year with Eddington and another year in the U.S. observatories. I visited Slipher and Hubble and heard him in Washington, in 1925, making his memorable communication about the distance [to] the Andromeda nebula. While my Mathematics bibliography was seriously in default since I did not know the work of Friedmann, it is perfectly up to date from the astronomical point of view; I calculate [in my contribution] the coefficient of expansion (575 km per sec per megaparsecs, 625 with a questionable statistical correction). Of course, before the discovery and study of clusters of nebulae, there was no point to establish the Hubble law, but only to calculate its coefficient. *The title of my note leaves no doubt on my intentions:* A Universe with a constant mass and increasing radius as an explanation of the radial velocity of extra-galactic nebulae. *I apologize that all of this is too personal*. But, as noted by the author (p. 161) "the history of this science competition is not irrevelant" and it is useful to highlight the details to enable an exact understanding of the scope of the argument that can be drawn from this. (Emphasis added)
>
> In 1950, Lemaître clearly did not want the rich fusion of theory and observations contained in his 1927 paper to be buried in the sands of time.